\documentclass[prx, superscriptaddress]{revtex4-1}
\usepackage{epsfig}
\usepackage{verbatim}
\usepackage{graphicx}   
\usepackage{dcolumn}    
\usepackage{bm}         
\usepackage{amssymb,amsmath}

\usepackage{color}
\usepackage{hyperref}
\hypersetup{colorlinks, citecolor=blue}

\begin{document}

\title{Automated Real-Space Lattice Extraction for Atomic Force Microscopy Images}

\author{Marco Corrias}
\affiliation{University of Vienna, Faculty of Physics and Center for
Computational Materials Science, Kolingasse 14-16, A-1090 Vienna, Austria}
\email{marco.corrias@univie.ac.at}
\affiliation{University of Vienna, Vienna Doctoral School in Physics,
Boltzmanngasse 5, 1090 Vienna, Austria}
\author{Lorenzo Papa}
\affiliation{University of Vienna, Faculty of Physics and Center for
Computational Materials Science, Kolingasse 14-16, A-1090 Vienna, Austria}
\author{Igor Sokolovi{\'c}}
\affiliation{Institute of Applied Physics, TU Wien, Wiedner Hauptstrasse 8-10/134, A 1040 Vienna, Austria}
\author{Viktor Birschitzky}
\affiliation{University of Vienna, Faculty of Physics and Center for
Computational Materials Science, Kolingasse 14-16, A-1090 Vienna, Austria}
\affiliation{University of Vienna, Vienna Doctoral School in Physics,
Boltzmanngasse 5, 1090 Vienna, Austria}
\author{Alexander Gorfer}
\affiliation{University of Vienna, Faculty of Physics and Center for
Computational Materials Science, Kolingasse 14-16, A-1090 Vienna, Austria}
\author{Martin Setvin}
\affiliation{Department of Surface and Plasma Science, Faculty of Mathematics and Physics, Charles University, V Holešovičkách 2, 180 00 Prague 8, Czech Republic}
\affiliation{Institute of Applied Physics, TU Wien, Wiedner Hauptstrasse 8-10/134, A 1040 Vienna, Austria}
\author{Michael Schmid}
\affiliation{Institute of Applied Physics, TU Wien, Wiedner Hauptstrasse 8-10/134, A 1040 Vienna, Austria}
\author{Ulrike Diebold}
\affiliation{Institute of Applied Physics, TU Wien, Wiedner Hauptstrasse 8-10/134, A 1040 Vienna, Austria}
\author{Michele Reticcioli}
\affiliation{University of Vienna, Faculty of Physics and Center for
Computational Materials Science, Kolingasse 14-16, A-1090 Vienna, Austria}
\author{Cesare Franchini}
\affiliation{University of Vienna, Faculty of Physics and Center for
Computational Materials Science, Kolingasse 14-16, A-1090 Vienna, Austria}
\affiliation{Dipartimento di Fisica e Astronomia, Universit\`{a} di Bologna, 40127 Bologna, Italy}

\begin{abstract}
\textbf{Abstract.}
Analyzing atomically resolved images is a time-consuming process requiring solid experience and substantial human intervention. In addition, the acquired images contain a large amount of information such as crystal structure, presence and distribution of defects, and formation of domains, which need to be resolved to understand a material's surface structure. Therefore, machine learning techniques have been applied in scanning probe and electron microscopies during the last years, aiming for automatized and efficient image analysis. This work introduces a free and open source tool (AiSurf: Automated Identification of Surface Images) developed to inspect atomically resolved images via Scale-Invariant Feature Transform (SIFT) and Clustering Algorithms (CA).
AiSurf extracts primitive lattice vectors, unit cells, and structural distortions from the original image, with no pre-assumption on the lattice and minimal user intervention.
The method is applied to various atomically resolved non-contact atomic force microscopy (AFM) images of selected surfaces with different levels of complexity: anatase TiO$_2$(101), oxygen deficient rutile TiO$_2$(110) with and without CO adsorbates, SrTiO$_3$(001) with Sr vacancies and graphene with C vacancies. The code delivers excellent results and has proved to be robust against atom misclassification and noise, thereby facilitating the interpretation scanning probe microscopy images.
\end{abstract}

\maketitle

\section{Introduction}
\label{sec:01}
In recent years giant leaps have been made in scanning probe microscopy, particularly in atomic force microscopy (AFM)~\cite{Giessibl2003Advances,morita2015noncontact,pavlivcek2017generation}.
Atomically sharp tips, often functionalized with simple molecules (e.g., CO-terminated tips)~\cite{scheuerer2019charge,kempkes2019design,fournier2011force,Wagner2012Measurement} allow not only for the structural identification of material surfaces at the atomic level but also for precise manipulation of single atoms and molecules~\cite{gross2009chemical,alldritt2020automated}. \\
Improved imaging techniques, aided by artificial intelligence (AI)~\cite{krull2020artificial, Kalinin2021a,Kalinin2022}, lead to a vast abundance of atomically resolved images.
Acquisition rates have indeed reached a point where the analysis of individual images by humans, albeit computer-assisted, is proving to be a new bottleneck in the advancement of related surface science fields. 

Machine learning (ML) can represent a viable alternative to accelerate the processing of these data. Supervised learning approaches based on deep neural networks (DNN) have been applied to a variety of different tasks~\cite{Choudhary2022}, ranging from tip functionalization~\cite{rashidi2018autonomous,gordon2020embedding} and lattice recognition~\cite{ziatdinov2018deep, kalinin2021bayesian, Ziletti2018, Leitherer2021}, all the way up to automated molecular structure discovery~\cite{alldritt2020automated,nano11071658}. 
Despite these promising results, the applicability of DNN for efficient interpretation of experimental images is limited by the available datasets~\cite{Kalinin2016}. While electron microscopy (EM)~\cite{Aversa2018, Schwenker2020, Ede2020} and scanning tunneling microscopy (STM)~\cite{Choudhary2019} image datasets have been recently published, we are not aware of publicly available dataset of atomically resolved experimental AFM images for surfaces (a database on simulated AFM images for molecular identification has been recently collected~\cite{doi:10.1021/acs.jcim.1c01323}).
Augmenting an image dataset using measured and simulated data are sub-optimal options for practical purposes:
(i) although AI optimization methods have been proposed~\cite{Kalinin2021a}, it is complicated to measure high-quality images and define a reliable reproducibility protocol;
(ii) simulating atomically resolved images using computational surface science methods requires the precise knowledge of the structural model at the atomic scale, which might be accessible for bulk-terminated surfaces but highly complex in the case of surface structural reconstructions; the vast diversity of defects that can be present in a surface complicates this task even further;
(iii) last but not least, supervised machine learning also heavily relies on the network architecture, which forbids non-expert users to play with different architectures to achieve better accuracies. Open-source packages like pycroscopy~\cite{somnath2019pycro} and AtomAI~\cite{ziatdinov2021atomai} can overcome this initial difficulty, although they remain dependent on a dataset.
These reasons hinder the adoption of supervised machine learning in the field of AFM microscopy.

In the absence of large datasets, unsupervised machine learning methods constitute a convenient alternative to data-hungry supervised ML techniques and slow manual analysis of individual images.
Several successful attempts based on unsupervised algorithms have been made.
Atomap~\cite{nord2017Atomap} analyzes scanning transmission electron microscopy (STEM) images by using 2D Gaussian fitting, differentiating atomic columns by their different brightness; different approaches based on clustering and principal component analysis (PCA) were also proposed~\cite{Belianinov2015, Somnath2018}.
In a recent work focused on the automated identification of local structures in atomically resolved images, Laanait \textit{et al.}~\cite{Laanait2016} have shown that the well-established computer vision algorithms  Scale-Invariant Feature Transform~\cite{Lowe2004} (SIFT) combined with Clustering Algorithms (CA)~\cite{gan2007data} are capable of recognizing and labeling atoms according to their local environment.
An advantage of this approach is that it does not involve any pre-assumptions on the underlying periodicity of the lattice. Reciprocal space analysis can be used to analyze periodic structures in images and, eventually, remove image artifacts. However, some prior knowledge of the system and the intervention by an expert user are required.
Restricting the analysis to real space is expected to provide better stability under such features, thereby expanding the degree of analysis.

In this work, we propose an open-source tool, AiSurf ("\textit{A}utomated \textit{i}dentification of \textit{Surf}ace images"), developed to inspect and classify crystalline 2D phases in atomically-resolved images via Scale Invariant Feature Transform and Clustering Algorithms.
With no pre-assumption on the lattice symmetry and only minimal human intervention, AiSurf can be applied for the analysis of regular structures, as well as for images showing complex structural correlations involving the formation of structural domains, structural and chemical defects, and recognition of adsorbates.
To assess the performance and transferability of the code, we analyze a variety of surface structures characterized by different features, specifically:
(i) Defect-free anatase TiO$_2$(101);
(ii) SrO-terminated SrTiO$_3$(001) with Sr vacancies;
(iii) Rutile TiO$_2$(110) with O vacancies;
(iv) CO adsorbates on rutile TiO$_2$(110);
(v) Simulated graphene with C vacancies.
Additionally, to further test the capabilities of the algorithm against experimental noise, we show the results obtained for an SrTiO$_3$(001) image 
affected by noise and other artifacts.
All AFM images have been acquired via AFM experiments, apart from the graphene image, which has been simulated using the probe particle model~\cite{hapala2014mechanism, hapala2016mapping}.

The algorithm and computational protocol are presented in the next section. The experimental details on the acquisition of the AFM images and their automated analysis are presented and discussed in Sec.~\ref{sec:results}. Data, code and documentation are available as indicated in Sec.~\ref{sec:data}.

\section{Method}
\label{sec:method}

\begin{figure}
    \centering
    \includegraphics[width=1\columnwidth]{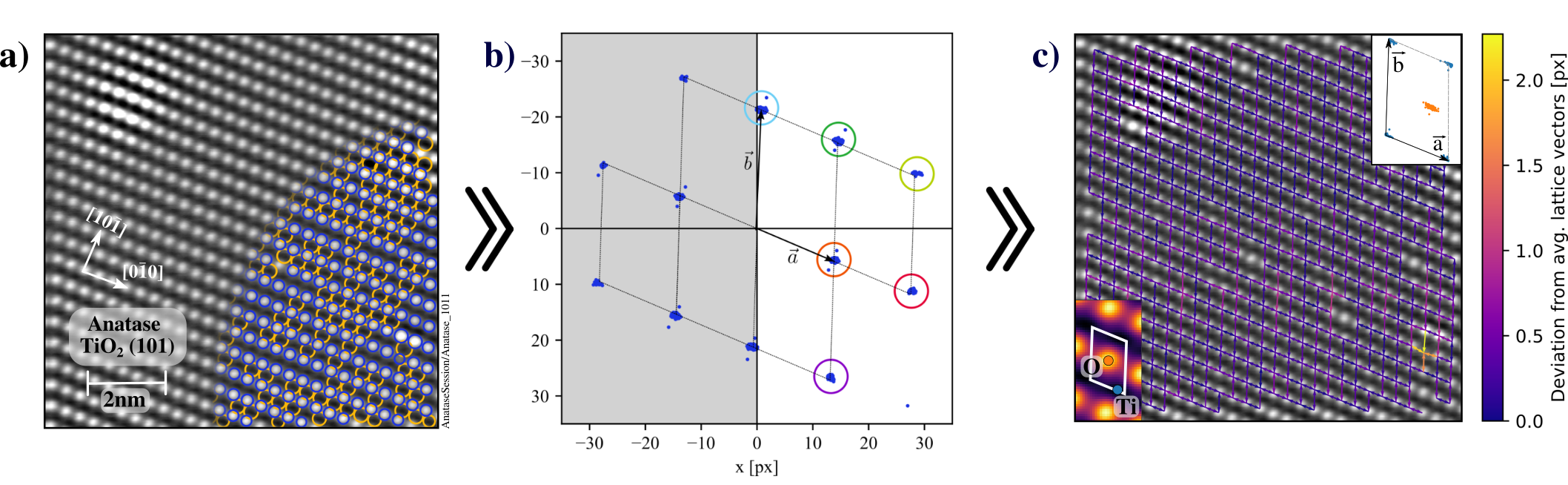}
    \caption{Summary of the workflow of AiSurf using an atomic force microscopy image of the anatase TiO$_2$(101) surface. a) AFM image with SIFT keypoints colored according to their cluster label. b) Distance vectors within the (blue) reference cluster. Colored circles show the result of a clustering done in this figure; each cluster represents a lattice vector candidate: $\Vec{a}$ and $\Vec{b}$ have been chosen. c) Distance vectors connecting keypoints. The colors of the vectors represent the deviation from the predicted lattice vectors; the top-right inset shows the distribution of keypoints in the unit cell, with keypoints colored according to their cluster; the bottom-left inset shows the predicted unit cell with its sublattice positions. We point out that what is interpreted as Ti atom is just a minimum, since Ti atoms are too far down to be reached by the tip.}
    \label{fig:methods}
\end{figure}

This section describes the workflow of AiSurf, graphically schematized in Fig.~\hyperref[fig:methods]{\ref{fig:methods}}.

First, we apply SIFT (as implemented in the library 'opencv'~\cite{opencv_library}) to an input image to extract atomic features ("keypoints") as bright and dark blobs, in analogy with the work of Laanait \textit{et al.}~\cite{Laanait2016}.
We specify that "atomic features" may also refer to adsorbates on the surface or, in general, to minima or maxima present in the image.
SIFT detects keypoints using the difference-of-Gaussians (DoG) function with sub-pixel resolution.
Although, for optimal detection, the input parameters have to be adjusted (depending on the scale and contrast of the features in the image), AiSurf default values can handle the images satisfactorily.
Every parameter is described in the documentation provided in Sec.~\ref{sec:data}. Ideally, all visible features should be identified while no keypoints should be assigned to background noise. To further enhance the quality of keypoints, we exclude a thin area on the border of the image, and we ignore keypoints showing a significant deviation in size from the median value.
This constraint should ensure that only atomically-sized features remain.

The keypoints detected by SIFT are characterized by a feature description based on image gradients that are brightness, scale, and rotational invariant.
We break rotational invariance to distinguish keypoints belonging to different sublattices if they show the same local environment but different orientations.
This can be the case of atoms of the same species but in different sublattices (e.g., graphene lattice, see Fig.~\hyperref[fig:results]{\ref{fig:results}}), or atoms next to a point defect (i.e. vacancy) but in different orientations.

The keypoints are then clustered based on their SIFT descriptors by using agglomerative clustering taken from the scikit-learn library~\cite{scikit-learn}.
This method is deterministic and only uses the number of clusters as an input parameter.
Density-based clusterings are challenging to use here because they are typically parameterized by a characteristic distance, which is difficult to estimate for abstract descriptors like ours, which are a collection of local gradients.
The optimal number of clusters is chosen by calculating the silhouette score~\cite{ROUSSEEUW198753} for each number of clusters in a chosen interval and taking the one that maximizes it.
A label is assigned to each cluster.

As an example, we show the results obtained for a defect-free AFM image of the anatase TiO$_2$(101) surface.
The experimental image overlaid with the clustered keypoints is shown in Fig.~\hyperref[fig:methods]{\ref{fig:methods}a}.
The clustering labels allow us to select a reference cluster that will be used to extract the lattice vectors.
It should contain only keypoints on the same sublattice since they have similar local environments and, thus, similar SIFT descriptors.
We point out that AiSurf is currently not designed to distinguish different domains; for example, translational or rotational domains~\cite{hommrich2002domains, lee2006domains} will be recognized as a single one. 
The reference cluster is by default the one with most keypoints, but users can change it at their preference.
In order to obtain the lattice vectors, the nearest neighbours are computed for every keypoint of the reference cluster.
These nearest neighbors are grouped into sub-clusters, used to calculate the distance vectors from the reference point.
The centers of mass of these sub-clusters represent possible lattice vector candidates.
The two shortest, linear independent candidates are chosen as lattice vectors $\Vec{a}$ and $\Vec{b}$.
Finally, a Bravais lattice is generated by linear combinations of these lattice vectors.
Fig.~\hyperref[fig:methods]{\ref{fig:methods}b} shows a visual representation of this process.
We note that the identification of lattice vectors and Bravais lattice is based uniquely on the reference cluster, and all other clusters are not verified to hold the same vectors:
this choice allows to exclude spurious effects arising from defects.

After determining the lattice vectors, it is possible to calculate the sublattice positions by processing all detected keypoints again.
For each keypoint we compute its position within the newly-obtained unit cell; the resulting distribution of positions within the unit cell is then clustered.
The center of mass of each cluster is calculated using the Bai-Breen method~\cite{bai2008calculating}, accounting for the periodic boundary conditions.
The resulting distribution of keypoints in the unit cell is visible in the top-right inset of Fig.~\hyperref[fig:methods]{\ref{fig:methods}c}. Up to this point, the keypoints were labeled according to the clustering performed on the descriptors, which allowed for the selection of the reference cluster for the lattice vectors identification. Now, every keypoint is labeled according to its sublattice position.

The algorithm can also plot an average view of the unit cell by averaging the area around each keypoint belonging to the same atomic species. The predicted unit cell and its sublattice positions are then drawn.
The bottom left of Fig.~\hyperref[fig:methods]{\ref{fig:methods}c} shows the image extracted with this method.

\section{Results and discussion}
\label{sec:results}

\subsection{Experimental and simulated AFM images}
The images of anatase TiO$_2$(101) in Fig.~\hyperref[fig:methods]{\ref{fig:methods}}, the SrO termination of cleaved SrTiO$_3$(001) in Fig.~\hyperref[fig:results]{\ref{fig:results}b,c} and rutile TiO$_2$(110) in Fig.~\hyperref[fig:application]{\ref{fig:application}} have been obtained using non-contact atomic force microscopy (nc-AFM)~\cite{morita2015noncontact,giessibl2019qplus} in constant-height mode and ultra-high vacuum, whereas the nc-AFM image of graphene was simulated. The signal displayed as grayscale image is the frequency shift $\Delta$f. Details on the experimental and computational setups are given below.

Fig.~\hyperref[fig:results]{\ref{fig:results}a} shows a constant-height nc-AFM image simulated using the probe particle model~\cite{hapala2014mechanism, hapala2016mapping} on free-standing graphene with C vacancies in a regular pattern, and three N atoms substituting C in the sites neighboring the vacancy. The simulation was performed using a CO-terminated AFM tip~\cite{gross2009chemical} (oriented with the O atom towards the surface) at a distance of 0.36\,nm from the most-protruding atom in the surface slab, using an oscillation amplitude of 200\,pm. The CO tip, with $-0.05e^-$ charge on the protruding O atom, was characterized by a spring constant of 0.5\,N/m. The simulation was performed over a $8\times8$\,nm$^2$ surface area, with $802\times 802$ pixels. 

Figure~\ref{fig:results}b shows an unreconstructed SrO-terminated region on a cleaved SrTiO$_3$(001) surface~\cite{sokolovic2019incipient, sokolovic2021quest}. This termination consists of Sr and O atoms, organized in a (1$\times$1) square pattern, with a characteristic concentration of 0.14$\pm$0.02 monolayers of point defects in the form of Sr vacancies. 
The image 
was acquired  in close vicinity to the surface using an O-terminated tip~\cite{sokolovic2020resolving}, such that Sr and O atoms are detected in the attractive (dark) and repulsive regime (bright), respectively; Sr vacancies are imaged as bright, cross-shaped features.
The gradually decaying contrast from the bottom to the top of the image is due to the over-compensation of vertical drift caused by creep of the piezo scanner, moving the AFM sensor away from the surface. 
The image was obtained with no application of tip-sample bias voltage, with an oscillation amplitude of 100\,pm, over a $5.4\times 5.4$\,nm$^2$ surface region with $400 \times 400$ pixels. Imaging was performed in ultra-high vacuum with base pressure below 1$\times$10$^{-11}$\,mbar, at a sample temperature of 5\,K. 

Fig.~\hyperref[fig:results]{\ref{fig:results}c} shows the same surface as Fig.~\hyperref[fig:results]{\ref{fig:results}b}.
In addition to the intrinsic Sr vacancies imaged in the form of cross-shaped point defects, this surface also hosts protruding defects (in a concentration of less than 0.5\%, displayed as dark regions) that typically appear after a surface is exposed to the residual gas in ultra-high vacuum for several days. The contrast differences in the four image regions are due to different tip-sample distances throughout the image since the nc-AFM tip was manually retracted/approached to the surface to avoid contact with the protruding defects. The image was obtained with an oscillation amplitude of 300 pm, without applying a bias voltage, over a 16.5$\times$16.5~nm$^2$ region with 400$\times$400~pixels. Imaging was performed under the same conditions of Fig.~\hyperref[fig:results]{\ref{fig:results}b}.

The nc-AFM image of clean rutile TiO$_2$(110) shown in Fig.~\hyperref[fig:application]{\ref{fig:application}a} was acquired during a previous study~\cite{reticcioli2017polaron}.
This surface consists of rows of two-fold coordinated bridging O$_{2c}$ atoms and five-fold coordinated Ti$_{5c}$ atoms, running along the $[001]$ and alternating along the $[1\bar{1}0]$ direction; the rows of O atoms are occasionally interrupted by single point defects in the form of oxygen vacancies.  
Bright spots indicate O atoms detected in the repulsive regime, whereas O vacancies are imaged as missing spots
(the Ti atoms are too distant from the tip and are not detected by nc-AFM).

The constant-height nc-AFM image of a rutile TiO$_2$(110) covered by CO molecules shown in Fig.~\hyperref[fig:application]{\ref{fig:application}b} was acquired during a previous study~\cite{Reticcioli2019Interplay}.
Imaging was performed with a CO-terminated tip that detects each adsorbed CO molecule as a bright spot; bridging oxygen atoms are not resolved because the adsorbed CO molecules protrude significantly more.

\subsection{Automated analysis and discussion}
\begin{figure*}[t]
    \centering
    \includegraphics[width=\textwidth]{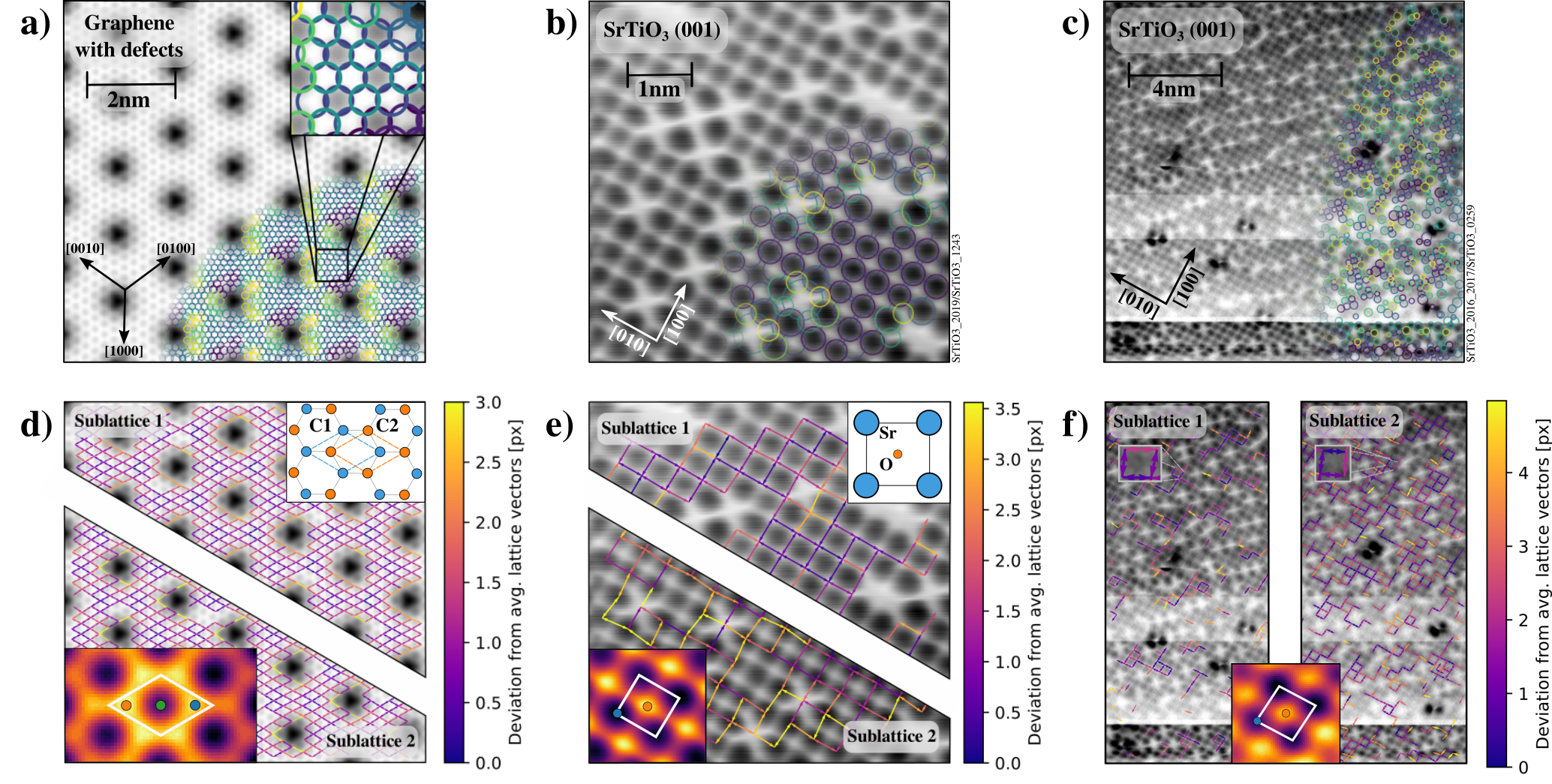}
    \caption{Application of AiSurf to different atomic force microscopy images: simulated graphene with C vacancies (a,~d); SrTiO$_3$(001) with Sr vacancies (b, c, e, f).
    Panels a--c: Surfaces with detected keypoints; color coding indicates different clusters. 
    On graphene, C atoms appear bright (repulsive).
    On SrTiO$_3$, Sr vacancies are imaged as bright, cross-shaped features, Sr atoms are imaged as dark spots, O atoms as bright spots.
    Panels d--f: Corresponding lattice analysis for different sublattices (the color gradients indicates deviations from the predicted perfect lattice); the bottom insets show the predicted unit cell.
    The top-right panels in (d) and (e) are schematic views of the surface structures.
}
    \label{fig:results}
\end{figure*}
The results obtained by AiSurf for graphene and SrTiO$_3$(001) are collected in Fig.~\hyperref[fig:results]{\ref{fig:results}}. The parameters used for this analysis are included in the code repository.

Three images were analyzed to test the capabilities of AiSurf under different conditions. Fig.~\hyperref[fig:results]{\ref{fig:results}a} shows a simulated image of graphene, used here to inspect the capability of our 
algorithm 
in fully controlled conditions; Fig.~\hyperref[fig:results]{\ref{fig:results}b} shows a typical experimental AFM image of SrTiO$_3$(001) with high contrast, which represents a more realistic case; the experimental image in Fig.~\hyperref[fig:results]{\ref{fig:results}c} shows multiple artifacts, and strong contrast variations;
it has been used to test the capabilities of the algorithm under challenging conditions.

Figs.~\hyperref[fig:results]{\ref{fig:results}a--c} show (part of) the keypoints detected by AiSurf, marked with coloured circles (each cluster with a different color). These are the keypoints left after the filtering process described in Sect.~\ref{sec:method}. It can be noticed how the SIFT algorithm can accurately detect the centers of the features, even if they are only a few pixels in size, as the ones in Fig.~\hyperref[fig:results]{\ref{fig:results}a,~c}. This is vital for the lattice recognition analysis since off-centered keypoints lead to inaccurate predictions. This first clustering process may not correctly detect all the different atomic species in the image but guarantees that the largest clusters contain features of the same type, which is needed for a successful analysis. Indeed, Figs.~\hyperref[fig:results]{\ref{fig:results}a--c} show that atoms surrounding defects are typically assigned to different clusters for the same atomic species far from the defects; this apparent misclassification gets corrected during the sublattice recognition, where keypoints are labelled according to their sublattice position instead of the descriptor.

\begin{figure*}[t]
    \centering
    \includegraphics[width=0.8\columnwidth]{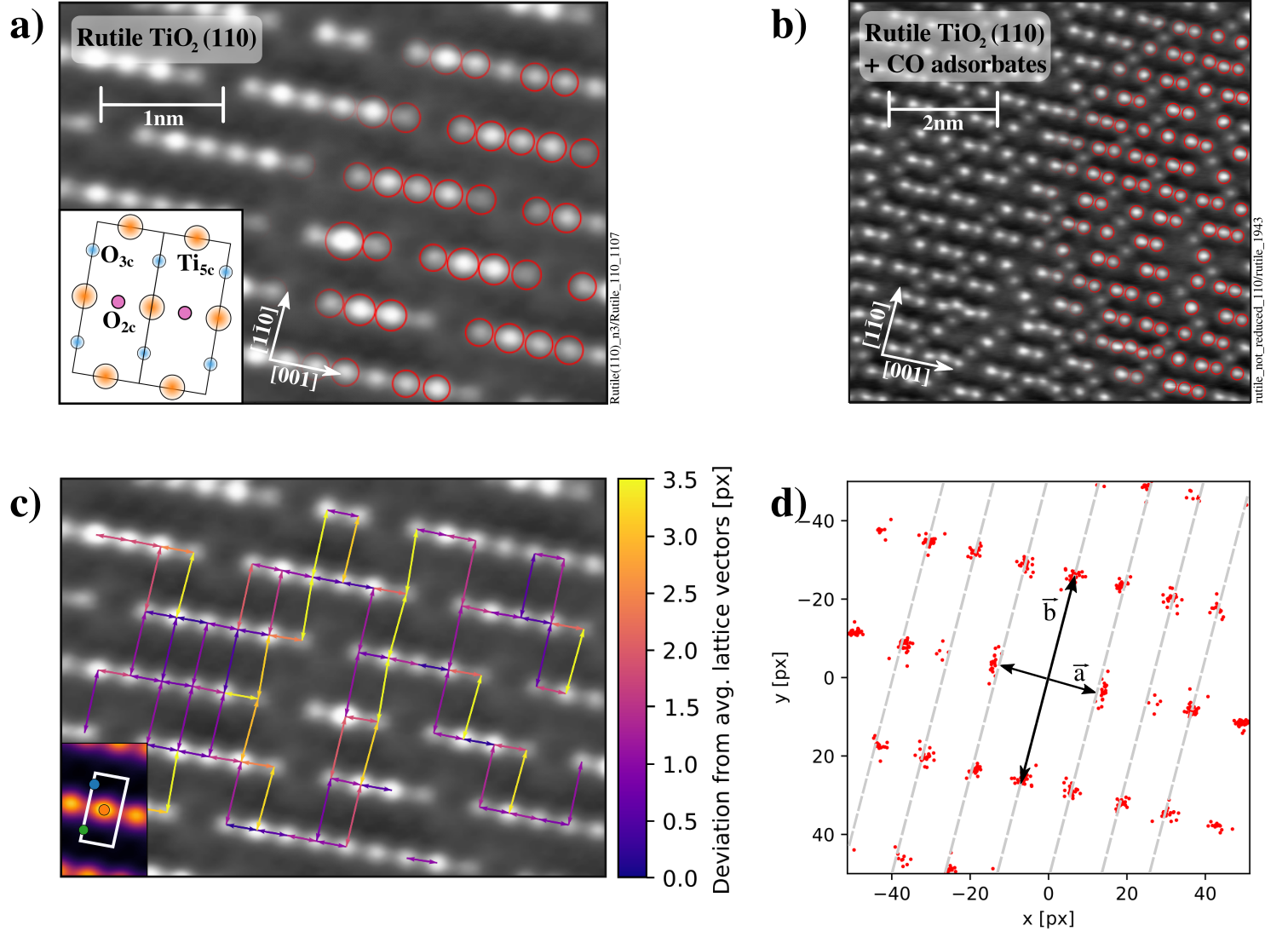}
    \caption{Automated analysis of experimental atomic force microscopy images of rutile TiO$_2$(110) with (b,~d) and without (a,~c) CO adsorbates.
    Panel a: bridging oxygen atoms protruding from the rutile surface show rows of bright spots interrupted by oxygen vacancies; the red circles indicates the automatic detection of these oxygen atoms. The inset shows a sketch of the surface structure: the protruding bridging oxygen atoms are labelled as 'O$_{2c}$', while the surface Ti and O atoms are labelled as 'Ti$_{5c}$' and 'O$_{3c}$', respectively. AFM image adapted from Ref.~\cite{reticcioli2017polaron}.
    Panel c: deviation plot, showing the deviations from the lattice vectors.
    The predicted unit cell is shown in the inset.
    Panel b: adsorbed CO molecules are shown as bright spots; CO on Ti$_{5c}$ atoms are marked with red circles; AFM image adapted from Ref.~\cite{Reticcioli2019Interplay}.
    Panel d: analysis of the CO-CO distance for every CO on Ti$_{5c}$ atom detected in the experimental image.
    }
    \label{fig:application}
\end{figure*}

In all cases, the identification of lattice vectors and unit cells works with good accuracy.
Since defects are neglected in the lattice vector extraction, they pose no challenge to the algorithm. However, some atoms are not detected, leading to gaps in the lattices shown in the bottom row images. In addition, the breaking of rotational invariance leads to a separation of the carbon sublattices (see Fig.~\hyperref[fig:results]{\ref{fig:results}a}) and a distinction between keypoints next to a defect (based on the direction the defect lies).
The knowledge of the lattice periodicity allows us to identify any deviation from a regular pattern. Fig.~\hyperref[fig:results]{\ref{fig:results}d--f} show the vectors between keypoints belonging to the same sublattice, with a color coding depicting their deviation from the predicted average lengths. 
This distance-deviation plot helps detect displacements due to the presence of defects and other sources. Fig.~\hyperref[fig:results]{\ref{fig:results}d} shows slight displacements only around the nitrogen impurities.
The top-right inset depicts the ideal lattice and highlights the two inequivalent C atoms ("C1" and "C2"). The bottom-left inset shows the average predicted unit cell; an orange and a blue circles represent the two carbon atoms. The algorithm has recognized the green circle at the center of the cell as a feature due to its strong contrast with the other ones, but it is known that such feature is an atom for hexagonal lattices, not honeycomb ones like this case. All the images, except for the insets with white background, have been automatically generated by the algorithm. \\
Fig.~\hyperref[fig:results]{\ref{fig:results}e} shows the Sr and O sublattices and their distance deviations. As expected, deviations are present in the proximity of defects; 'Sublattice 2' shows some misclassified points but evident distortions. The square unit cell has been flawlessly predicted. Fig.~\hyperref[fig:results]{\ref{fig:results}f}, used to test the algorithm's capabilities, shows overall positive results. Sublattice 1, defined by Sr atoms, presents several unrecognized areas, especially in the low-contrast regions. Distance deviations might not be reliable in this case. Sublattice 2 posed a lower challenge for its detection. The zoomed areas show that the algorithm can detect features even when barely visible, proving its high robustness against noise. The unit cell has been successfully predicted.

Fig.~\hyperref[fig:application]{\ref{fig:application}} shows the application of AiSurf to rutile TiO$_2$(110) with and without CO adsorbates~\cite{Reticcioli2019Interplay}. Fig.~\hyperref[fig:application]{\ref{fig:application}a} shows some detected bridging oxygen atoms, highlighted in red. 
The corresponding deviation plot is shown on Fig.~\hyperref[fig:application]{\ref{fig:application}c}: distance deviations near oxygen vacancies are visible; away from vacancy sites, no relevant deviations are present. This image is a good example of how defect-induced deviations can be easily highlighted with AiSurf. Such local structural distortions could play functional role in chemical reactions between surface and adsorbates and might also facilitate the formation and identification of metastable configurations~\cite{meier2022}. The inset in Fig.~\hyperref[fig:application]{\ref{fig:application}c} shows the average unit cell. Other than the central oxygen atom (labeled as  "O$_{2c}$" in Fig.~\hyperref[fig:application]{\ref{fig:application}a} inset), two other features are present in the cell, interpreted as three-fold coordinated oxygen atoms ("O$_{3c}$") given their position with respect to O$_{2c}$.

Figs.~\hyperref[fig:application]{\ref{fig:application}b,d} show the analysis of CO molecule adsorbed on rutile TiO$_2$(110). CO molecules adsorb both on five-fold coordinated Ti atoms and oxygen vacancies. CO on Ti atoms are highlighted with red circles in Fig.~\hyperref[fig:application]{\ref{fig:application}b}. The CO-CO distances distribution on these sites is shown in panel (d).
In the latter one can easily visualize details barely visible from the original AFM image:
neighboring CO along $\vec{a}$ ([001] direction) tend to repel each other by slightly tilting away from the surface normal, unlike second nearest neighbor CO or CO belonging to different rows. This can be noticed by observing the different shape of the two clusters at the center of panel (d), which are differently distributed than the others, marking an absence of short-distance neighboring CO.
This result agrees with density functional theory calculations~\cite{Kunat2009CO, Prates2017CO}. Without computer tools, such details can be overlooked in content-rich images like Fig.~\hyperref[fig:application]{\ref{fig:application}b}. \\

While AiSurf is an aid for finding features in atomically resolved AFM images, determination of unit cells, and deviations from a regular arrangement, the interpretation is mostly left to the user. First, it has to be noted that the SIFT algorithm is based on detection of local minima and maxima. A minimum or maximum is not necessarily a physical feature, as exemplified by the hollow sites in the hexagonal graphene rings if Fig.~\hyperref[fig:results]{\ref{fig:results}a,d} or the minima in Fig.~\hyperref[fig:methods]{\ref{fig:methods}c} (blue dot in the bottom inset, no atom is detected in this position, Ti atom is an interpretation). In the SrTiO$_3$ case, the maxima between the dark Sr sites (sublattice 1 in Fig.~\hyperref[fig:results]{\ref{fig:results}e,f}) are at the positions of the O atoms, but this is a mere coincidence. Areas with Sr vacancies show that the O sublattice is not resolved; these areas appear with roughly constant brightness and do not show the O atoms as maxima.

When analyzing displacements, it is important to be aware of the influence of the tip on imaging. Especially tips with a rather wobbly termination such as CO-functionalized tips can easily deform due to (electrostatic) forces on the tip~\cite{hapala2014mechanism}; this can lead to apparent displacements of the surface atoms~\cite{hapala2016mapping}. Oxygen-terminated tips tend to be stiffer, but even in the absence of tip deformation the minima or maxima in $\Delta$f images may not exactly coincide with the atomic positions. In our SrTiO$_3$ example, the Sr vacancies are charged defects, 
thus they will distort the electrostatic field in their vicinity. Since the negative frequency shift above the Sr atoms is due to electrostatic attraction between the O-terminated tip and the positive Sr atom, the field of a neighboring vacancy will influence the position of the $\Delta$f minimum. Thus, the displacements around the Sr vacancies (yellow and orange vectors in Fig.~\hyperref[fig:results]{\ref{fig:results}e}) do not necessarily reflect the atomic coordinates, and they may be also related to the imaging process. \\

\section{Summary and Conclusions}
In this work, various atomically resolved non-contact AFM images with different surface symmetries and degrees of complexity were analyzed in an automated way using the open source tool AiSurf.
At its present state, AiSurf allows detecting the distribution of interatomic distances in distinct sublattices and extracts primitive lattice vectors, unit cell, and distance deviations from ideally symmetric lattices. We have shown results obtained on simulated graphene with vacancies, as well as experimental images of anatase and rutile TiO$_2$ (the latter with and without CO adsorbates) and SrTiO$_3$(001) with Sr vacancies. The distance-deviation plot proved to be useful for the detection of lattice distortions caused by oxygen vacancies, and the CO tilting (see Fig.~\hyperref[fig:application]{\ref{fig:application}c,~d} respectively). This could be applied to the analysis of any sources of bond deviation and similiar.
The algorithm performed well even on a SrTiO$_3$(001) experimental image strongly affected by noise and artifacts. Such robustness suggests a future implementation for real-time analysis, parallel to the image acquisition process.
For this reason, the application on atomically resolved images other than AFM ones (e.g. scanning tunneling microscopy) is believed to be possible and will be further tested. In addition, future developments are already planned, such as the identification and analysis of grain boundaries, domains and detection of defects.

AiSurf is a user-friendly, documented unsupervised-ML tool that requires minimal user intervention and no need to provide any image database.
Developing a robust tool for analyzing atomically resolved images is still an ambitious goal for both supervised and unsupervised techniques. The most adopted ones deliver successful results for limited types of images, which usually present a good atomic contrast and/or are acquired with a precise technique, such as the one for which they were designed.
We aim to design a protocol that relies on little or no prior knowledge of the physical system, hoping to achieve a wide degree of flexibility. 
To conclude, we believe that providing intuitive and open access tools will help material scientists accelerate time-consuming tasks like image selection and analysis, and assist the detection of elusive features in atomically-resolved images.

\section{Data availability statement}
\label{sec:data}
The AiSurf tool along with the necessary input data will be soon publicly available.

\section*{Acknowledgments}
This work was supported by the Austrian Science
Fund (FWF) project Super (Grant No. P 32148-N36) and the SFB F81 project TACO. The computational results have been achieved using the Vienna Scientific Cluster (VSC).
MSe acknowledges the support from the Czech Science Foundation
GACR 20-21727X.
The code has been originally developed by L. Papa, and subsequently optimized by M. Corrias.

\bibliography{references}

\end{document}